\pdfoutput=1
\documentclass[aps,pra,preprint,nofootinbib,superscriptaddress, floatfix,groupedaddress]{revtex4-1}
\usepackage[english]{babel}
\usepackage[utf8]{inputenc}
\usepackage{mathtools}
\usepackage{braket}
\usepackage{lineno}
\usepackage{pdfpages}
\usepackage{stfloats}
\usepackage{natbib}
\usepackage{xcolor}
\usepackage[symbol]{footmisc}
\makeatletter
\AtBeginDocument{\let\LS@rot\@undefined}
\makeatother

\usepackage[justification=raggedright,singlelinecheck=false]{caption,subcaption}
\usepackage{siunitx}

\usepackage{scalerel}
\usepackage{tikz}
\usetikzlibrary{svg.path}
\definecolor{orcidlogocol}{HTML}{A6CE39}
\tikzset{
  orcidlogo/.pic={
    \fill[orcidlogocol] svg{M256,128c0,70.7-57.3,128-128,128C57.3,256,0,198.7,0,128C0,57.3,57.3,0,128,0C198.7,0,256,57.3,256,128z};
    \fill[white] svg{M86.3,186.2H70.9V79.1h15.4v48.4V186.2z}
                 svg{M108.9,79.1h41.6c39.6,0,57,28.3,57,53.6c0,27.5-21.5,53.6-56.8,53.6h-41.8V79.1z M124.3,172.4h24.5c34.9,0,42.9-26.5,42.9-39.7c0-21.5-13.7-39.7-43.7-39.7h-23.7V172.4z}
                 svg{M88.7,56.8c0,5.5-4.5,10.1-10.1,10.1c-5.6,0-10.1-4.6-10.1-10.1c0-5.6,4.5-10.1,10.1-10.1C84.2,46.7,88.7,51.3,88.7,56.8z};}}
\newcommand\orcidicon[1]{\href{https://orcid.org/#1}{\mbox{\scalerel*{
\begin{tikzpicture}[yscale=-1,transform shape]
\pic{orcidlogo};
\end{tikzpicture}
}{|}}}}
\usepackage[colorlinks,citecolor=red,urlcolor=blue,bookmarks=false,hypertexnames=true]{hyperref}

\begin{document}

\title{Two-phonon scattering in non-polar semiconductors: a first-principles study of warm electron transport in Si}

\author{Benjamin Hatanp{\"a}{\"a} \orcidicon{0000-0002-8441-0183}}
\affiliation{Division of Engineering and Applied Science, California Institute of Technology, Pasadena, CA, USA}

\author{Alexander Y. Choi \orcidicon{0000-0003-2006-168X}}
\affiliation{Division of Engineering and Applied Science, California Institute of Technology, Pasadena, CA, USA}

\author{Peishi S. Cheng \orcidicon{0000-0002-3513-9972}}

\author{Austin J. Minnich \orcidicon{0000-0002-9671-9540}}
\thanks{Corresponding author: \href{mailto:aminnich@caltech.edu}{aminnich@caltech.edu}}

\affiliation{Division of Engineering and Applied Science, California Institute of Technology, Pasadena, CA, USA}

\date{\today} 

\begin{abstract}
The \textit{ab-initio} theory of charge transport in semiconductors typically employs the lowest-order perturbation theory in which electrons interact with one phonon (1ph). This theory is accepted to be adequate to explain the low-field mobility of non-polar semiconductors but has not been tested extensively beyond the low-field regime. Here, we report first-principles calculations of the electric field-dependence of the electron mobility of Si as described by the warm electron coefficient, $\beta$. Although the 1ph theory overestimates the low-field mobility by only around 20\%, it overestimates $\beta$ by over a factor of two over a range of temperatures and crystallographic axes. We show that the discrepancy in $\beta$ is reconciled by inclusion of on-shell iterated 2-phonon (2ph) scattering processes, indicating that scattering from higher-order electron-phonon interactions is non-negligible even in non-polar semiconductors. Further, a $\sim 20$\% underestimate of the low-field mobility with 2ph scattering suggests that non-trivial cancellations may occur in the perturbative expansion of the electron-phonon interaction. 
\end{abstract}

\maketitle

\noindent
\textit{Introduction.--}Recent advances in the first-principles treatment of charge transport in semiconductors have enabled the calculation of the low-field electrical mobility without any adjustable parameters \cite{Bernardi_2016,Giustino_2017}. The method, based on Wannier interpolation of electron-phonon matrix elements \cite{mostofi_2008,Pizzi_2020}, allows the Boltzmann equation to be solved on a sufficiently fine grid to ensure converged transport properties. Calculations of low-field mobility have been reported for various semiconductors, including Si \cite{Li_2015,Fiorentini_2016,Ponce_2018}, GaAs \cite{Zhou_2016,Liu_2017,Lee_2020}, and others \cite{Ma_2018,Sun_2015,Lee_2018}. Methodological developments continue to be reported, including an ab-initio treatment of two-phonon scattering \cite{Lee_2020} and the quadrupole electron-phonon interaction \cite{Park_2020}. A recent work has extended these methods to magnetotransport \cite{Desai_2021}, high-field transport \cite{Maliyov_2021}, and transport and noise of warm and hot electrons in GaAs \cite{Choi_2021, Cheng_2022} and holes in Si \cite{Catherall_2022}.

The accuracy of first-principles theory has been tested primarily by computing the low-field mobility and comparing to the available experimental data at various temperatures and doping concentrations. The warm electron regime, defined as the regime in which the next-to-leading order term of the expansion of current density with electric field is non-negligible, is of interest because it contains information on the band structure anisotropy \cite{Schmidt_1963} and energy relaxation \cite{Costato_1973} not evident in the low-field mobility. The non-Ohmic mobility of Ge and Si beyond the low-field regime was first reported by Shockley \cite{Shockley_1951} and Ryder \cite{Ryder_1953}. Subsequent investigation led to the prediction \cite{Shibuya_1955} and experimental observation \cite{Sasaki_1958, Jorgensen_1963,Gibbs_1964,Hamaguchi_1966,Jorgensen_1967} of the anisotropy of the mobility at high electric field in multi-valley semiconductors owing to the differential heating of transverse and longitudinal valleys, known as the Sasaki-Shibuya effect. In 1963, Schmidt-Tiedemann reported a theory of the warm electron tensor, showing that in cubic crystals the fourth-rank warm electron tensor can be completely described by two independent components owing to crystal symmetry \cite{Schmidt_1963}. The two independent components are denoted $\beta$ and $\gamma$, with $\beta$ describing the variation of conductivity with electric field and $\gamma$  the non-parallelism of the current and electric field. Substantial experimental data versus temperature and crystallographic direction is available for both $\beta$ and $\gamma$ for electrons in Si \cite{Brown_1961,Kastner_1965,Gibbs_1964,Jorgensen_1963, Hamaguchi_1963}.

Although scattering by the interaction of an electron with one phonon (1ph) has typically been employed in theoretical and Monte Carlo studies at low field and in the warm electron regime to interpret transport studies, \cite{conwell_1967,Jorgensen_1967,Hamaguchi_1963,Asche_1969,Fawcett_1970} other experiments suggest a non-negligible role for higher-order processes \cite{Dumke_1960,Thomas_1968,Stradling_1968,Folland_1970}.  In Si, two-phonon (2ph) deformation potentials were extracted from second-order Raman spectra \cite{Temple_1973,Renucci_1975}, and calculations of charge transport properties based on these values have indicated that 2ph scattering may make a  non-negligible contribution to scattering rates \cite{Kubakaddi_1977,Ngai_1974, Alldredge_1967}. Recent ab-initio works have reported that two-phonon scattering plays a role in both low-field and high-field transport in the polar semiconductor GaAs \cite{Lee_2020,Cheng_2022}. Despite these works, the accepted conclusion from ab-initio studies is that 1ph scattering is sufficient to describe the low-field mobility of non-polar semiconductors \cite{Ma_2018}. However, this conclusion has not been extensively tested away from the low-field regime.

Here, we report first-principles calculations of the warm electron tensor in Si. At the 1ph level of theory, both the low-field mobility and $\beta$ are overestimated, with a marked discrepancy of $\beta$ at 300 K of over a factor of two. To address this discrepancy, we compute the scattering due to sequential 1ph processes, corresponding to one of the terms at second-order in the electron-phonon interaction. The scattering rates are found to be comparable to those of 1ph scattering over a range of energies, and their inclusion eliminates the discrepancy in $\beta$. The resulting $\sim 20$\% underestimate of mobility suggests that accounting for cancellations of the two second-order terms of the electron-phonon interaction may be necessary to achieve quantitative agreement for both the mobility and $\beta$.

\noindent
\textit{Theory and numerical methods.--} We begin by describing the numerical approach to compute the transport properties of electrons in Si in the warm electron regime. The low-field mobility calculation is now routine and has been extensively described in Refs.~\cite{Bernardi_2016,Giustino_2017}. In the low-field regime, the carrier temperature equals the lattice temperature and the current density varies linearly with the electric field. The proportionality coefficient is simply $\sigma_0$, the linear DC conductivity. 


The warm electron regime is defined by electric fields for which the cubic term of the expansion of current density with electric field becomes non-negligible \cite{Schmidt_1963}. Mathematically, the current density vector $\mathbf{j}$ in the warm electron regime can be expanded in powers of the electric field of magnitude $E$ as $j_i = E \sigma_0 e_i + E^3 \sigma_{iklm} e_k e_l e_m + ...$ where $e_i$ are the components of the unit vector in the direction of the electric field along Cartesian axis $i$ and $\sigma_{iklm}$ is the fourth-rank warm electron conductivity tensor. In cubic crystals, warm electron transport is fully defined by two components of the tensor, $\beta$ and $\gamma$, where $\sigma_{1111} = \sigma_0 \beta$ and $\sigma_{1122} = \frac{1}{3}\sigma_0 (\beta - \gamma)$. Equivalently, the warm electron tensor can be specified by the values of $\beta$ along different crystallographic axes.

The warm electron tensor has not been calculated previously by ab-initio methods. It may be obtained from the first-principles approach described in Sec.~2 of Ref.~\cite{Choi_2021} to solve the Boltzmann equation in the warm electron regime. Briefly, the Boltzmann equation for a non-degenerate, spatially homogeneous electron gas subject to an applied electric field $\mathcal{E}$ is given by Eq. 2 of Ref.~\cite{Choi_2021}. As described in this reference, in the warm electron regime the reciprocal space derivative must be applied to the total distribution function. In the present treatment, the full derivative is evaluated using a finite difference approximation given in \cite{mostofi_2008} and Eqn.~8 of Ref.~\cite{Pizzi_2020}. The Boltzmann equation then becomes a linear system of equations (Eqn.~5 in Ref.~\cite{Choi_2021}) that can be solved by numerical linear algebra methods. With the distribution function obtained, transport quantities can be obtained by appropriate Brillouin zone integrations.

For the present study of electrons in Si, the electronic structure and electron-phonon matrix elements are computed on a coarse $8 \times 8 \times 8$ grid using DFT and DFPT with \textsc{Quantum Espresso} \cite{Giannozzi_2009}. We employ a plane wave cutoff of 40 Ryd and a relaxed lattice parameter of 5.431 $\textrm{\AA}$. We set the Fermi level 203 meV below the conduction band minimum (CBM) corresponding to a non-degenerate electron gas of concentration of $10^{16}$ cm${}^{-3}$ at 300 K. The energy window of the Brillouin zone was set to 287 meV above the CBM. In \textsc{Perturbo}, a grid of $100^3$ for the Wannier interpolated electronic structures and electron-phonon matrix elements was employed. Increasing the energy window to 447 meV changed the mobility by 0.1\% and $\beta$ by 0.8\%, while increasing the grid density to $140^3$ resulted in mobility changes on the order of 1\%. As in prior work, spin-orbit coupling is neglected as it has a weak effect on electron transport properties in Si \cite{Ponce_2018,Ma_2018}. Quadrupole electron-phonon interactions were neglected as they provide only a small correction to the low-field mobility of silicon at room temperature \cite{Park_2020}.

\begin{figure}[h]
{\phantomsubcaption\label{mobility_vs_temperature}
\phantomsubcaption\label{beta_vs_temperature}}
\includegraphics[width=\columnwidth]{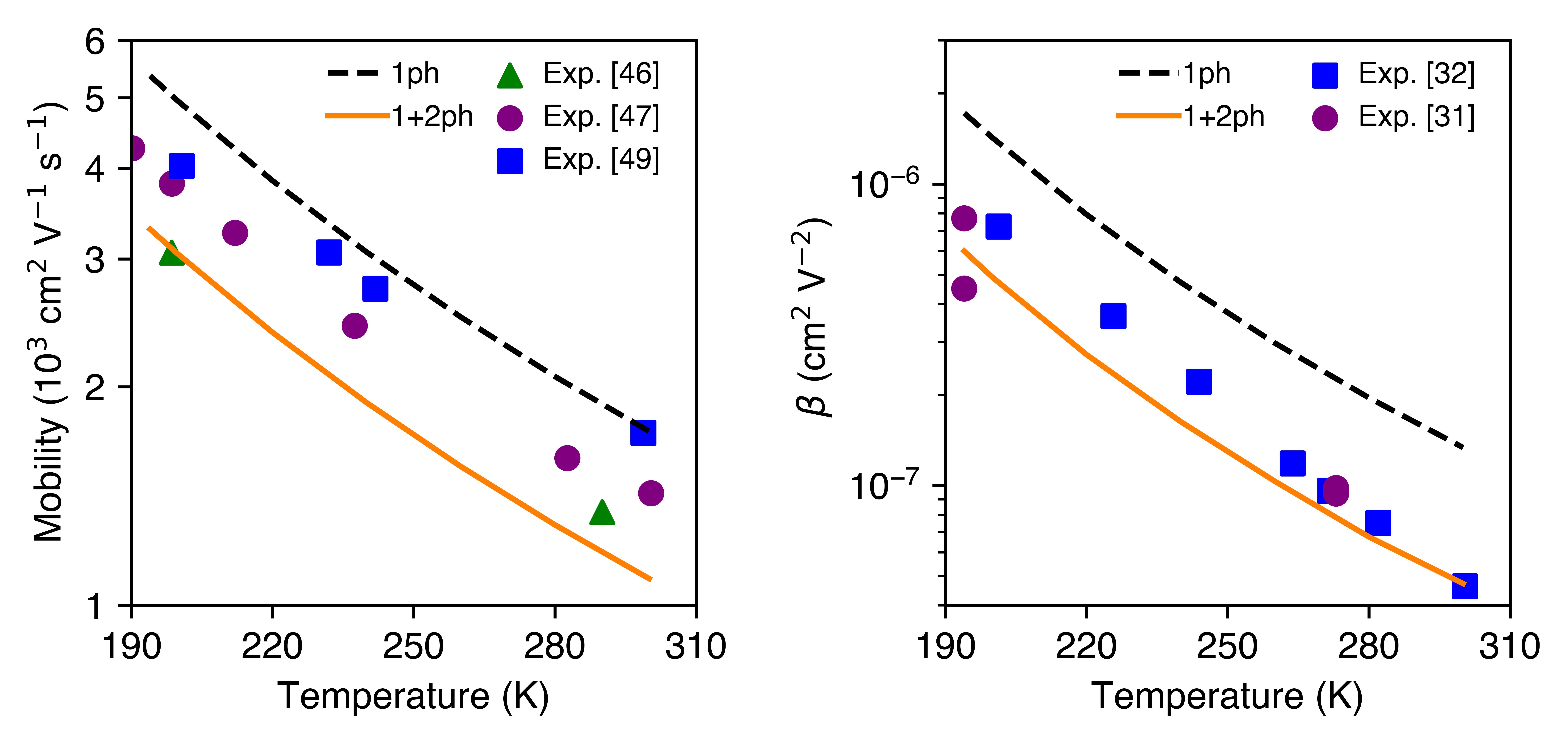}
\caption{(a) Low-field mobility versus temperature for the 1ph (dashed black line) and 1+2ph (solid orange line) frameworks. The mobility is overestimated with the 1ph level of theory, but underestimated when including on-shell 2ph scattering. Experimental data: Fig.~11, Ref. \cite{Canali_1975} (green triangles), Fig.~1, Ref. \cite{Norton_1973} (purple circles), Fig.~2, Ref. \cite{Logan_1960} (blue squares). (b) $\beta$ versus temperature for the 1ph (dashed black line) and on-shell 2ph (solid orange line) frameworks. $\beta$ is overestimated by $\gtrsim 100$\%  at the 1ph level of theory across all temperatures. When including on-shell 2ph, the discrepancy is eliminated. Experimental data from Fig.~11, Ref. \cite{Hamaguchi_1963} (blue squares) and Figs.~3 and 4, Ref. \cite{Kastner_1965} (purple circles).}
\label{temperature}
\end{figure}

\noindent
\textit{Mobility and warm electron tensor.--} Figure \ref{mobility_vs_temperature} shows the computed low-field mobility versus temperature for electrons in Si at the 1ph level of theory. The low-field value at 300 K is 1737 $\text{cm}^2 \text{V}^{-1} \text{s}^{-1}$, approximately 20\% higher than experimental values, which range between 1300 and 1450 $\text{cm}^2 \text{V}^{-1} \text{s}^{-1}$ \cite{Canali_1975,Norton_1973,Sze_2006}. The calculated value at 300 K is generally consistent with prior ab-initio studies, which report values of 1915 $\text{cm}^2 \text{V}^{-1} \text{s}^{-1}$  \cite{Ma_2018}, 1860 $\text{cm}^2 \text{V}^{-1} \text{s}^{-1}$  \cite{Li_2015}, and 1750 $\text{cm}^2 \text{V}^{-1} \text{s}^{-1}$ \cite{Fiorentini_2016}. The use of the experimental lattice constant and GW quasiparticle corrections leads to lower mobility values \cite{Ponce_2018}. The general overestimate of low-field mobility in past ab-initio studies applies across a wide range of temperatures, as Refs.~\cite{Li_2015} and \cite{Ma_2018} show an overestimated mobility from 100 to 300 K.

At higher fields, the mobility decreases below the low-field value owing to electron heating. The warm electron coefficient $\beta$ can be obtained by fitting the quadratic dependence of the mobility on electric field. We define the fitting range as that corresponding to a 0.1 percent increase in electron temperature, although the value of $\beta$ is not particularly sensitive to this choice. Because $\beta$ depends on the direction of the applied field, we denote $\beta$ without subscripts to indicate the electric field is applied along the [100] crystallographic axis.

\begin{figure}[h]
\includegraphics[width=\columnwidth]{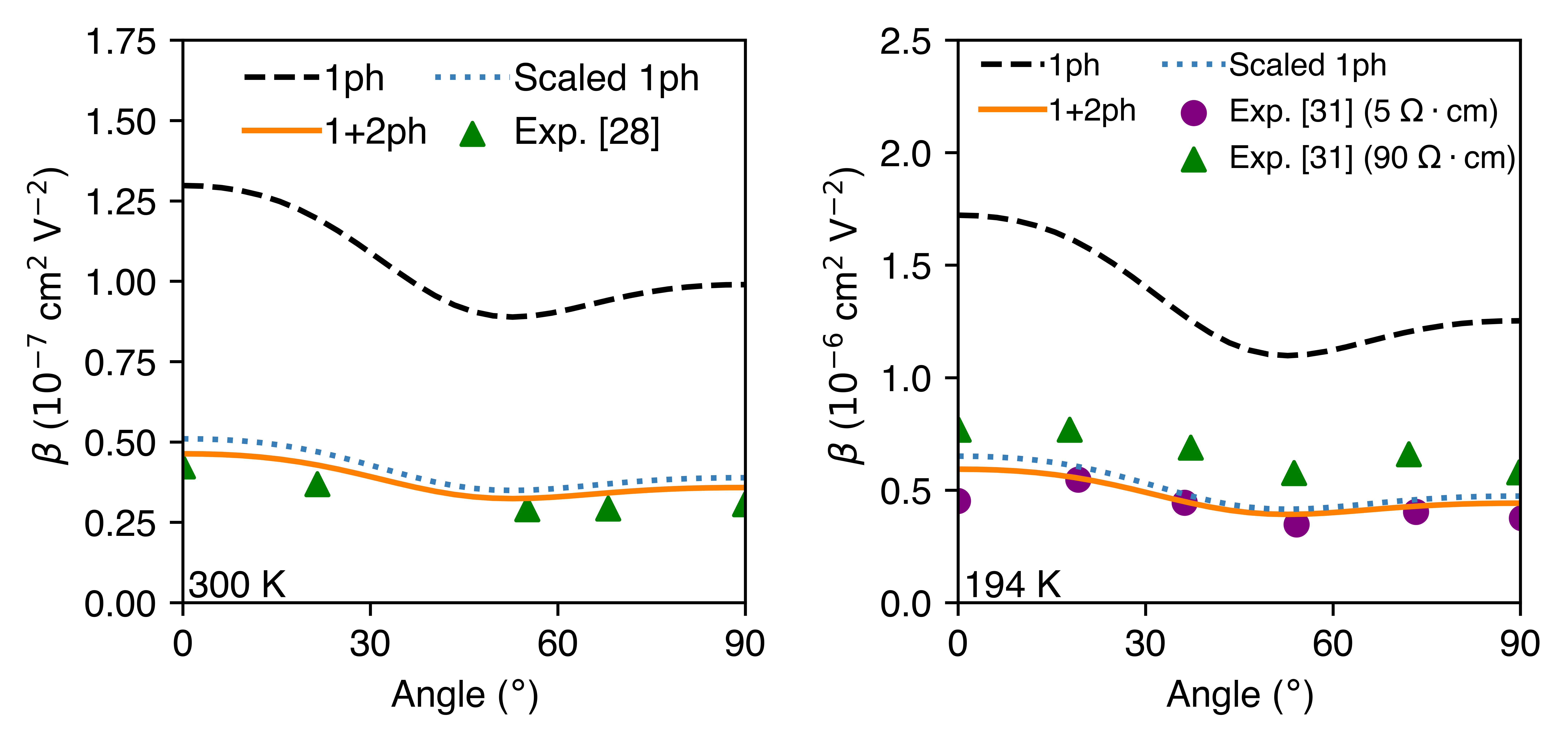}
{\phantomsubcaption\label{beta_vs_angle_300}
\phantomsubcaption\label{beta_vs_angle_194}}
\caption{\label{beta_vs_angle}(a) $\beta$ versus electric field orientation angle between the [001] and [110] crystallographic axes at 300 K for the 1ph (dashed black line), scaled 1ph (dotted blue line), and 1+2ph (solid orange line) frameworks. The 1ph theory captures the qualitative dependence of $\beta$ on angle, but the value is overestimated by $\sim$ 200\%. The discrepancy is reduced to $\sim 15$\% with inclusion of on-shell 2ph scattering. Experimental data from Fig.~7, Ref. \cite{Hamaguchi_1966} (upward green triangles). (b) Same as (a) at 194 K. Data from Figs.~3 and 4, Ref. \cite{Kastner_1965} (purple circles and green triangles).}
\end{figure}

In Figure \ref{beta_vs_temperature} we compare the computed $\beta$ versus temperature with experimental data. The prediction from the 1ph level of theory is clearly larger than the experimental values by around 150\% across all temperatures. This discrepancy is markedly larger than that of the low-field mobility. Figures \ref{beta_vs_angle_300} and \ref{beta_vs_angle_194} show $\beta$ versus angle between current direction and electric field at 300 and 194 K, respectively. Here, the mobility is presented as the field is rotated from the [001] direction (corresponding to 0$^{\circ}$) to the $[110]$ direction (corresponding to 90$^{\circ}$). While the qualitative trend in $\beta$ with field orientation seen in experiment is captured by the 1ph theory, the computed results again are greater than experiment by over $100\%$ at both temperatures.

\noindent
\textit{Role of higher-order phonon scattering.--} Figures \ref{beta_vs_temperature} and \ref{beta_vs_angle} indicate that $\beta$ is markedly overestimated at the 1ph level of theory. The magnitude of discrepancy  cannot be easily explained by inaccuracies in band structure, as the discrepancies of the effective mass are $\sim 7$\%. Therefore, we considered whether higher-order phonon scattering processes could account for the poor agreement. The 1ph level of theory accounts for the leading-order electron-phonon scattering process for which electrons scatter with one phonon. We implemented a treatment of the next-to-leading order scattering processes where electrons scatter with two phonons using the ab-initio approach described in Ref.~\cite{Lee_2020}. As in Ref.~\cite{Cheng_2022}, beyond the low-field regime the full 2ph calculation is presently computationally intractable, and so we included only on-shell 2ph processes that are within 25 meV of a band energy. Despite the neglect of off-shell processes, Ref.~\cite{Cheng_2022} indicates that most of the relevant processes are included with the approximation used here. The 2ph rates were iterated 5 times.

\begin{figure}[h]
{\phantomsubcaption\label{rates_1}
\phantomsubcaption\label{rates_2}}
\includegraphics[width=\columnwidth]{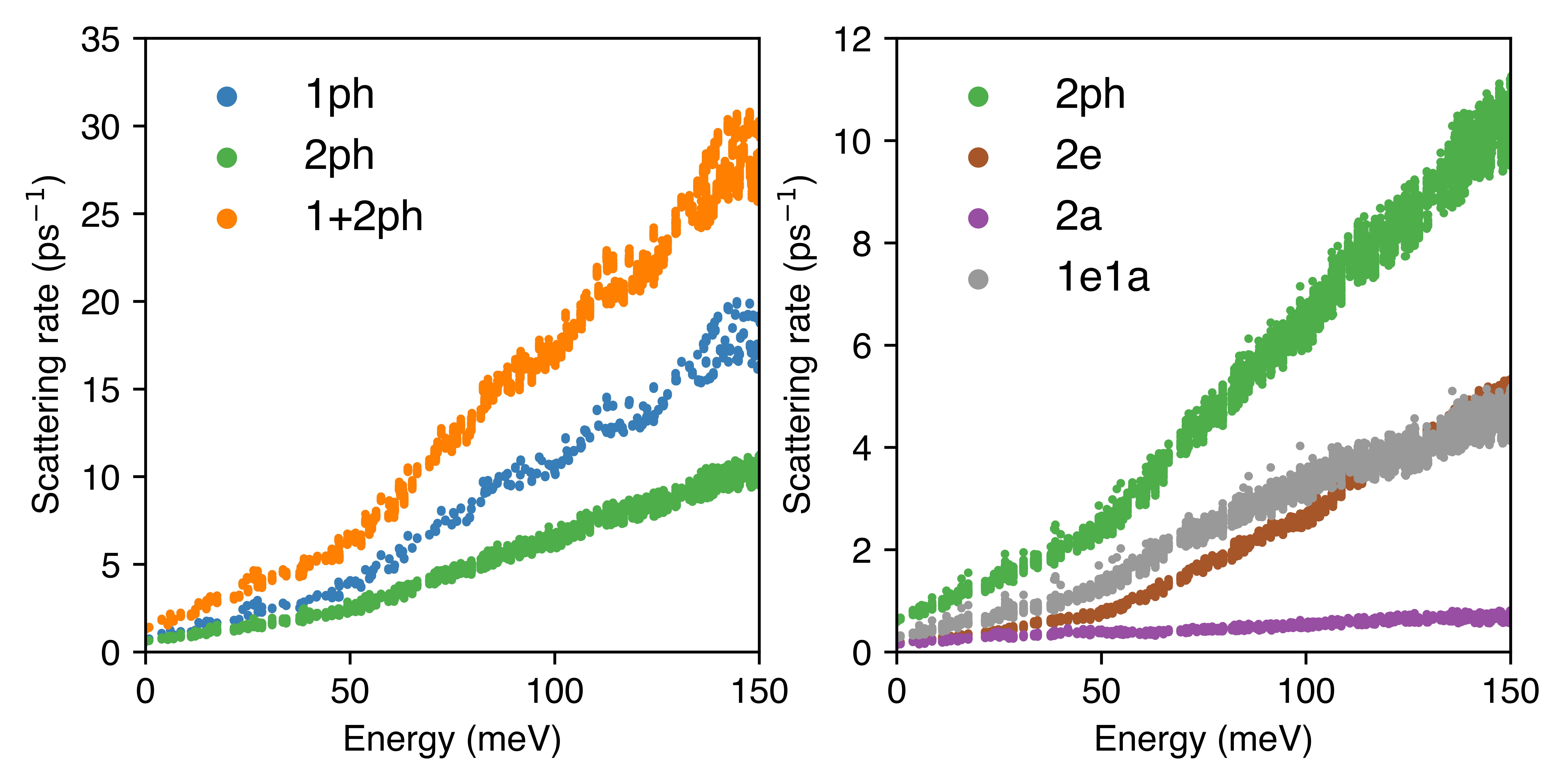}
\caption{\label{rates}(a) Computed 1ph (blue), on-shell 2ph (green), and 1+2ph (orange) scattering rates versus energy at 300 K. The on-shell 2ph rates are approximately 50\% of the 1ph rates, indicating a non-negligible contribution to transport properties. (b) Computed on-shell 2ph (green), 2e (brown), 2a (purple), and 1e1a (gray) scattering rates versus energy at 300 K. For energies less than 100 meV, the range relevant to transport properties at 300 K, the 1e1a rates are largest and have the dominant effect on transport properties.}
\end{figure}

The computed one and two-phonon scattering rates are shown in Fig.~\ref{rates_1}. Near the CBM, the 2ph rates are comparable to the 1ph rates. At 100 meV, the maximum energy relevant for transport properties at 300 K, the 2ph rate is approximately 50\% of the 1ph rate owing to the weaker energy dependence of 2ph scattering. A disaggregation of the rates into specific emission and absorption processes is shown in Fig.~\ref{rates_2}. For energies less than 100 meV, the 1e1a (one-phonon emission plus one-phonon absorption) rates are the largest and thus have the largest effect on transport properties, while the 2e (two-phonon emission) rates rise once electrons are able to emit two optical phonons. The 2a (two-phonon absorption) rates are relatively negligible at all energies and are only weakly dependent on energy. These characteristics are qualitatively similar to those reported for GaAs in Refs.~\cite{Lee_2020, Cheng_2022}.

We now examine the impact of the on-shell 2ph rates on the low-field mobility and $\beta$. In Figure \ref{mobility_vs_temperature}, the computed mobility versus temperature including on-shell 2ph scattering is shown. The computed 1+2ph curve underestimates experiment by about 20\%. At 300 K, the low-field mobility is 1089 $\text{cm}^2 \text{V}^{-1} \text{s}^{-1}$. In Figure \ref{beta_vs_temperature}, $\beta$ including on-shell 2ph scattering versus temperature is shown. With the inclusion of on-shell 2ph scattering, good agreement is observed with two independent experimental reports \cite{Hamaguchi_1963,Jorgensen_1967}. Similarly, in Figure \ref{beta_vs_angle_300}, the agreement with experiment \cite{Hamaguchi_1966} of the dependence of $\beta$ on orientation angle at 300 K is greatly improved by including on-shell 2ph. The qualitative trend of a decrease in $\beta$ from 0$^{\circ}$ until $\sim$ 55$^{\circ}$ (corresponding to the electric field in the [111] direction), followed by an increase until 90$^{\circ}$ is unchanged, but it is uniformly decreased in magnitude. The computed $\beta$ dependence on orientation angle at 194 K shown in Figure \ref{beta_vs_angle_194} lies between two data sets \cite{Kastner_1965} of different resistivities.

We now consider the origin of the the improved agreement with $\beta$ when including on-shell 2ph processes. The first mechanism is the increase in scattering rates, which have a relatively larger effect on $\beta$ compared to mobility. Specifically, it can be shown that for a uniform scaling of the scattering rates by a factor $\epsilon$, $\beta$ is scaled by $\epsilon^{-2}$ rather than $\epsilon^{-1}$ as for the mobility. Therefore, the increased scattering rates contributed by on-shell 2ph processes can account for part of the relatively larger decrease in $\beta$. To examine how much of the decrease in $\beta$ was due to the increased scattering rates, we scaled the 1ph scattering rates by a multiplicative factor so that the resulting low-field mobility was equal to the low-field mobility in the 1+2ph case. The DC mobility versus electric field and $\beta$ were then calculated. The scaled 1ph results can be seen in Figure \ref{beta_vs_angle}. At both 300 and 194 K, the majority (94\%) of the decrease in $\beta$ occurs due to the higher scattering rates. However, the calculated values of $\beta$ with the actual on-shell 2ph scattering rates are still lower than those predicted from the scaled 1ph rates. This further decrease is due to the larger sensitivity of $\beta$  to the scattering rate at low energies near the CBM compared to that of the mobility (see Supplementary Material). At these energies, 2ph processes make a relatively larger contribution to the scattering rates than at higher energies, leading to a larger reduction in $\beta$ than expected based on a uniform increase in scattering rates.

\noindent
\textit{Discussion.--}  We now discuss the finding that multi-phonon processes are relevant to transport in non-polar semiconductors. Previous experiment and modeling works have suggested that 2ph processes could account for deviations in the predicted temperature dependence of the mobility from the 1ph deformation potential theory. In particular, two-phonon deformation potentials were extracted from second-order Raman scattering measurements \cite{Temple_1973,Renucci_1975}, and using these values in transport calculations improved the agreement of both the variation of the low-field mobility and $\beta$ with temperature \cite{Kubakaddi_1977}. However, these conclusions were subject to uncertainty owing to the semi-empirical nature of the scattering rates employed in the modeling. The present work overcomes this limitation using the ab-initio scattering rates that are free of adjustable parameters, thereby providing firm evidence that multi-phonon scattering processes are of importance to low-field and warm electron transport in Si.

We additionally consider the role of other multi-phonon processes that have been neglected in the present study and their potential impact on the transport properties. First, the addition of the neglected off-shell 2ph processes will further increase the scattering rates and decrease both the mobility and $\beta$; however, Fig.~2a of Ref.~\cite{Cheng_2022} indicates this difference is negligible in GaAs. A more involved complication is the role of the direct 2ph interaction arising from simultaneous interactions with two phonons, in contrast to that arising from two sequential 1ph scattering events considered here. Due to translational invariance, a cancellation occurs for interactions involving long-wavelength acoustic phonons \cite{Holstein_1959}, and it has been posited that this cancellation may extend to acoustic phonons beyond this limit \cite{Kocevar_1980}. To estimate the magnitude of this cancellation, we removed all two-phonon processes that involve acoustic phonons of energy less than 5 meV and recalculated the low-field mobility. The result is 1261 $\text{cm}^2 \text{V}^{-1} \text{s}^{-1}$, which is in near-quantitative agreement with experiment. This result indicates that taking into account the cancellation between the two 2ph vertices may be needed for predictive accuracy. Further tests of the role of multi-phonon processes may be obtained by calculating the free carrier absorption spectrum using the methods of Ref. \cite{Noffsinger_2012} and the power spectral density of current fluctuations as in Refs. \cite{Choi_2021,Cheng_2022}.


\noindent
\textit{Summary.--} We have presented a first-principles calculation of the warm electron transport properties of Si. At the 1ph level of theory that is typically regarded as adequate for non-polar semiconductors, the low-field mobility is overestimated by around 20\% while $\beta$ is overestimated by over 100\% across a wide range of temperatures and crystallographic axes. The discrepancy in $\beta$ is reconciled by inclusion of 2ph scattering, which is found to exhibit a scattering rate that is comparable to that from 1ph processes. The  underestimate of the mobility at this level of theory provides evidence for the occurrence of a non-trivial cancellation of second-order terms in the electron-phonon interaction. 


\begin{acknowledgements}
B.H. was supported by a NASA Space Technology Graduate Research Opportunity. A.Y.C., P.S.C., and A.J.M were supported by AFOSR under Grant Number FA9550-19-1-0321. The authors thank J. ~Sun. S.~Sun, D.~Catherall and T.~Esho for helpful discussions.
\end{acknowledgements}

\bibliographystyle{is-unsrt}
\bibliography{references}{}

\clearpage

\begin{minipage}{\textwidth}
\includepdf[pages={1}]{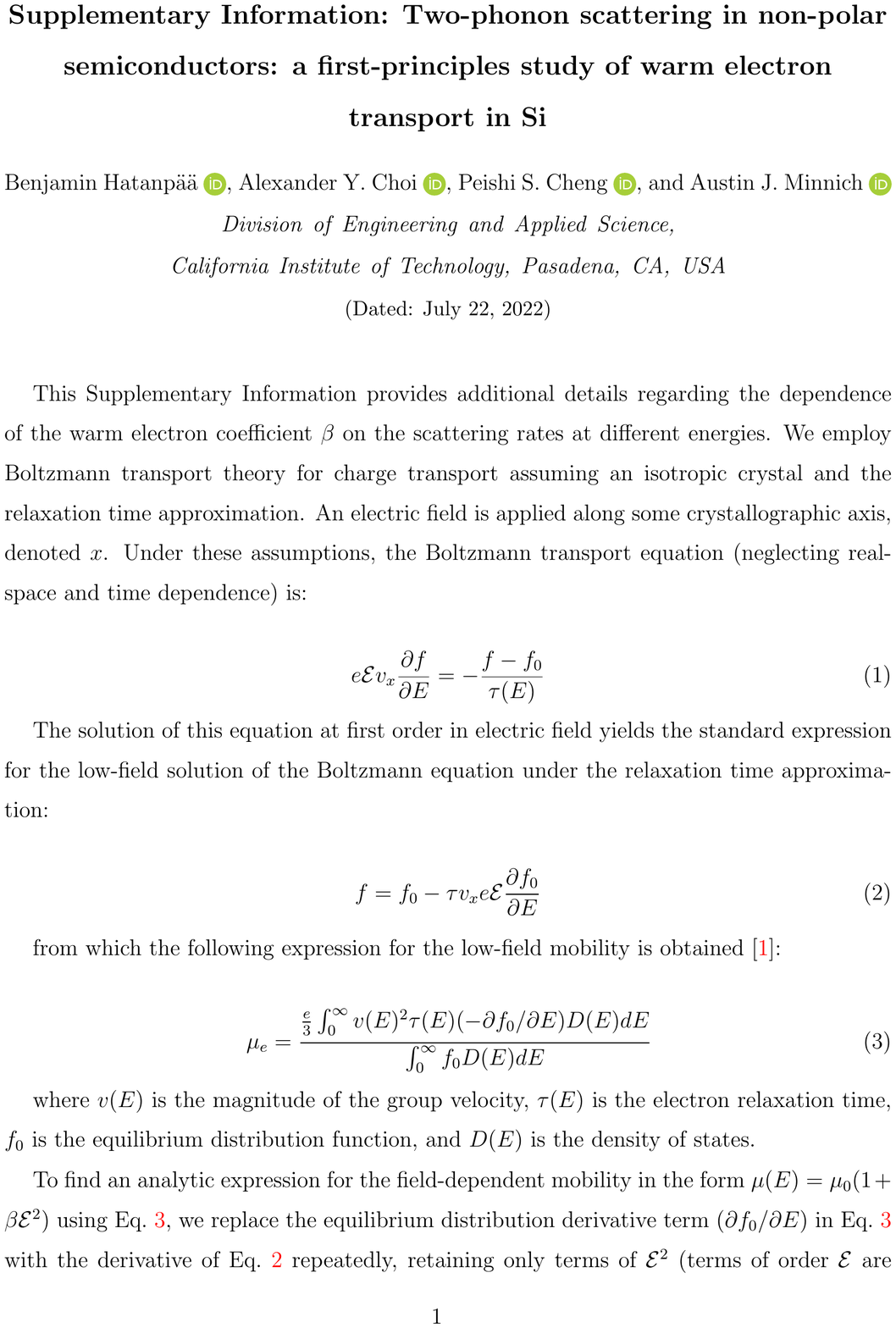}
\end{minipage}

\clearpage

\begin{minipage}{\textwidth}
\includepdf[pages={2}]{supplemental.pdf}
\end{minipage}

\clearpage

\begin{minipage}{\textwidth}


\section*{Supplementary material}

# Supplementary Information: Two-phonon scattering in non-polar semiconductors: a first-principles study of warm electron transport in Si

Benjamin Hatanpää [iD], Alexander Y. Choi [iD], Peishi S. Cheng [iD], and Austin J. Minnich [iD]

*Division of Engineering and Applied Science,*

*California Institute of Technology, Pasadena, CA, USA*

(Dated: July 22, 2022)

This Supplementary Information provides additional details regarding the dependence of the warm electron coefficient $\beta$ on the scattering rates at different energies. We employ Boltzmann transport theory for charge transport assuming an isotropic crystal and the relaxation time approximation. An electric field is applied along some crystallographic axis, denoted $x$. Under these assumptions, the Boltzmann transport equation (neglecting real-space and time dependence) is:

$$e\mathcal{E}v_x\frac{\partial f}{\partial E} = -\frac{f - f_0}{\tau(E)} \qquad (1)$$

The solution of this equation at first order in electric field yields the standard expression for the low-field solution of the Boltzmann equation under the relaxation time approximation:

$$f = f_0 - \tau v_x e\mathcal{E}\frac{\partial f_0}{\partial E} \qquad (2)$$

from which the following expression for the low-field mobility is obtained [1]:

$$\mu_e = \frac{\frac{e}{3}\int_0^\infty v(E)^2\tau(E)(-\partial f_0/\partial E)D(E)dE}{\int_0^\infty f_0 D(E)dE} \qquad (3)$$

where $v(E)$ is the magnitude of the group velocity, $\tau(E)$ is the electron relaxation time, $f_0$ is the equilibrium distribution function, and $D(E)$ is the density of states.

To find an analytic expression for the field-dependent mobility in the form $\mu(E) = \mu_0(1+\beta\mathcal{E}^2)$ using Eq. 3, we replace the equilibrium distribution derivative term $(\partial f_0/\partial E)$ in Eq. 3 with the derivative of Eq. 2 repeatedly, retaining only terms of $\mathcal{E}^2$ (terms of order $\mathcal{E}$ are

dropped as the final expression for electric current must be odd in electric field). In more detail, if we take the derivative of Eq. 2 with respect to $E$, we get

$$\frac{\partial f}{\partial E} = \frac{\partial f_0}{\partial E} + e\mathcal{E}\frac{\partial}{\partial E}\left(\tau v_x \frac{\partial f_0}{\partial E}\right) \tag{4}$$

Plugging in the expansion in Eq. 2 again for the second derivative term, we obtain

$$\frac{\partial f}{\partial E} = \frac{\partial f_0}{\partial E} + e\mathcal{E}\frac{\partial}{\partial E}\left(\tau v_x \frac{\partial f_0}{\partial E} + e\mathcal{E}\frac{\partial}{\partial E}\left(\tau v_x \frac{\partial f_0}{\partial E}\right)\right) \tag{5}$$

Discarding the terms of order $E$ due to symmetry, Eq. 5 becomes

$$\frac{\partial f}{\partial E} = \frac{\partial f_0}{\partial E} + (e\mathcal{E})^2 \frac{\partial^2}{\partial E^2}\left(\tau v_x \frac{\partial f_0}{\partial E}\right) \tag{6}$$

To simplify this expression, we assume that $f_0 \propto e^{-E/kT}$, $v_x \propto \sqrt{E}$ for a parabolic band, and $\tau \propto E^s$. The prefactors for these quantities are lumped into a constant, $c_0$. Plugging these expressions into Eq. 6, we obtain

$$\frac{\partial f}{\partial E} = \frac{\partial f_0}{\partial E} - \left(\frac{c_0}{kT}\right)(e\mathcal{E})^2 \frac{\partial^2}{\partial E^2}\left(E^{s+0.5}e^{-E/kT}\right) \tag{7}$$

Carrying out the second derivative with respect to energy $E$ in Eq. 7, we obtain

$$\frac{\partial f}{\partial E} = \frac{\partial f_0}{\partial E} - \left(\frac{c_0 e^{-E/kT}}{kT}\right) \times (e\mathcal{E})^2 \times \\ \left((s+0.5)(s-0.5)E^{s-1.5} - \left(\frac{2}{kT}\right)(s+0.5)E^{s-0.5} + \left(\frac{1}{kT}\right)^2 E^{s+0.5}\right) \tag{8}$$

We therefore identify an expression for $\beta$ up to a prefactor as

$$\beta \propto \int_0^\infty v(E)^2 \tau(E) e^{-E/kT} \times \\ \left((s+0.5)(s-0.5)E^{s-1.5} - \left(\frac{2}{kT}\right)(s+0.5)E^{s-0.5} + \left(\frac{1}{kT}\right)^2 E^{s+0.5}\right) D(E) dE \tag{9}$$

by the definition of mobility in Eq. 3 and gathering the terms at each order of electric field.

This expression can be simplified by substituting the energy dependencies of $v(E)$ and $D(E) \propto \sqrt{E}$ for a parabolic band and the assumed power law form for $\tau(E)$. For the mobility, we obtain

$$\mu_0 \propto \int_0^\infty E^{s+1.5} e^{-E/kT} dE \tag{10}$$

For $\beta$, we find

$$\beta \propto \int_0^\infty e^{-E/kT} \times$$
$$\left( (s+0.5)(s-0.5)E^{2s} - \left(\frac{2}{kT}\right)(s+0.5) E^{2s+1} + \left(\frac{1}{kT}\right)^2 E^{2s+2} \right) dE \tag{11}$$

Due to the terms with $E^{2s}$ and $E^{2s+1}$ in the expansion for $\beta$, the contribution to the integral from low energies is increased compared to that for the low-field mobility. For instance, for $s < 0$ as occurs for most semiconductors, the leading term in the integrand for $\beta$ will have a negative exponent, causing $\beta$ to depend on the low-energy scattering rates to a larger extent compared to the low-field mobility.

---


\end{minipage}

\end{document}